\documentclass[twocolumn,showpacs,amsmath,amssymb]{revtex4}

\usepackage{graphicx}
\usepackage{amsmath}
\usepackage{amssymb}

\begin{document}

\title{Magnetic adatoms on graphene in the Kondo regime:  an Anderson model treatment}
\author {Zhen-Gang Zhu,  and Jamal Berakdar}
\affiliation{ Institut f\"{u}r Physik
Martin-Luther-Universit\"{a}t Halle-Wittenberg
Nanotechnikum-Weinberg, Heinrich-Damerow-Strasse 4
D - 06120 Halle (Saale), Germany }

\begin{abstract}
We study theoretically the Kondo effect for a magnetic adatom on graphene using the Anderson model.
Upon obtaining the Green's function of the impurity to higher order contributions in the hybridization,
we calculated analytically the selfenergy in the presence  of  strong correlations.
It is found that the Kondo resonance takes place in a narrow energy range of the
impurity level around the Fermi energy which can be tuned by a gate voltage.
We show that this range is linear in the Fermi energy $|\mu|$ and is significantly  narrower than in the case  for a normal metal.
  The origin of this behavior is traced back to  the inherent properties of graphene, especially its linear dispersion.
  The singularity in the full Green's function is also analyzed with the help of a transparent geometrical method.
  The relations between the various selfenergies and the implications for the experimental observations are discussed .
\end{abstract}

\pacs{ 75.20.Hr, 73.22. Pr, 72.15.Qm, 81.05.ue, 71.20.Tx }
 \maketitle

\section{Introduction}
Recently, graphene, a two-dimensional monolayer of carbon atoms, has attracted much attention
\cite{graphene}. Its low energy band structure consists of two inequivalent Dirac cones located at the Brillouin zone
corners \cite{grapheneold}. In these cones, the energy-dispersion
relation is linear (relativistic), and the dynamics of the charge carriers is
governed by a massless Dirac-type equation.  The physics of magnetic adatoms on graphene is under active research by means  of the scanning tunneling spectroscopy  \cite{meyer,cornaglia,zhuang}, first-principles calculation \cite{chan}, and  many-body approaches that  deal with
the Kondo effect \cite{kotov,uchoa,ding,uchoa0906,hentschel,dora,sengupta,jacob,dellanna,vojta,zhu10,zhu2011,uchoa11}. Theoretical studies on magnetic impurities coupled to 2D Dirac fermions may be initiated in the context of \textit{d}-wave superconductors \cite{fradkin,buxton,fritz}. New aspects are particular to a graphene host such as the possibility of doping via a gate voltage that moves the Fermi level $\mu$ away from the Dirac point \cite{vojta}, and the different symmetries realized depending on where the adatom  is positioned  on the graphene lattice\cite{zhu10}.
Several studies have dealt with the physics of magnetic impurities on graphene.  A Hartree-Fock approximation \cite{uchoa,ding,uchoa0906} based on the Anderson model was utilized to study the mean-field behavior which is only valid at
$T>T_K$, where $T_K$ is the Kondo temperature. The Anderson model with infinite Coulomb correlation ($U$) under a magnetic field was studied
with the conclusion of a Fermi liquid behaviour \cite{dora}. On the other hand, the Kondo model (or \textit{sd} model) is also used to address spin interactions between the impurity spin and the spins of the Dirac fermions in graphene. The anisotropic \emph{single} channel Kondo model  \cite{hentschel} and a two-channel Kondo model \cite{sengupta} were both considered. However, for half spin, the former leads to a Fermi-liquid character; while the latter gives rise to a non-Fermi-liquid-like ground state. These two are essentially different \cite{bulla}. Very recently, the Kondo model was employed  to address the quantum criticality and  an expression of the Kondo temperature was given \cite{uchoa11}.  
In Ref. \cite{zhu10},  a detailed
symmetry group analysis was conducted to clarify the appropriate physical model. The conclusion is that,
   different symmetries are realized
 depending on whether the adatom is  above one carbon atom  (case A) or  on the center of the honeycomb (case B).
In Ref. \cite{zhu2011}, we focused on the Anderson model for the case A and conducted a Slave-Boson (SB) calculation. The results showed that the full occupation survives in a broad gate voltage range until the Fermi level tuned by the applied gate voltage touches the edge of the impurity level, in which case the zero-field susceptibility is finite. However,
the mean-field SB method does not answer how the Kondo effect evolves  with the temperature and other parameters. Therefore, to address these
 issues transparently, we calculate here the Green's function including higher orders in the hybridization by using the method developed in Ref. \cite{lacroix} and discuss how the graphene properties are reflected in the calculated physical quantities.
\section{Theoretical model}
The tight-binding Hamiltonian of the electrons in graphene is given
by
\begin{equation}
H_{\text{g}}=-t\sum\limits_{\langle i,j\rangle,\sigma}
(\bar{a}_{i,\sigma}^\dag \bar{b}_{j,\sigma}+\text{H.c.}), \label{hg}
\end{equation}
where $\bar{a}_{i,\sigma}^\dag (\bar{a}_{i,\sigma})$ creates
(annihilates) an electron with the spin $\sigma$ on the position
$\mathbf{R}_i$ of
 the sublattice A, $\bar{b}_{j,\sigma}^\dag (\bar{b}_{j,\sigma})$ creates
(annihilates) an electron with the spin $\sigma$ on the position
$\mathbf{R}_j$ on the sublattice B, $\langle i,j\rangle$ stands for
summation over nearest neighbors, and $t$ is the nearest neighbor
hopping energy. The Hamiltonian $H_{\text{g}}$ is
rewritten in the momentum space as
\begin{equation}
H_{\text{g}}=\sum\limits_{\mathbf{q},\sigma}
[\phi(\mathbf{q})\bar{a}_{\mathbf{q}\sigma}^\dag
\bar{b}_{\mathbf{q}\sigma}+\phi(\mathbf{q})^*
\bar{b}_{\mathbf{q}\sigma}^\dag \bar{a}_{\mathbf{q}\sigma}
],\label{hg1}
\end{equation}
where $\phi(\mathbf{q})=-t\sum\limits_{i=1}^3
e^{i\mathbf{q}\cdot\boldsymbol{\delta}_{i}}$. The Hamiltonian (\ref{hg1}) can be easily
diagonalized to give $E_{\pm}(\mathbf{q})=\pm
t|\phi(\mathbf{q})|$, which can be linearized around the
Dirac points in the Brillouin zone
\begin{equation}
E_{\pm}(\mathbf{k})=\pm v_F|\mathbf{k}|,
\end{equation}
where $v_{F}=3ta/2$ is the Fermi velocity of the electron ($t\sim 2.3$ eV
\cite{graphene}). Please, note $\mathbf{k}$ is now a momentum with respect to the Dirac points.

When introducing the impurity, we consider the situation where it is above one carbon atom of the graphene sheet. The symmetry for this case gives rise to a mixture of the two Dirac cones and the Hamiltonian reads \cite{zhu10}
\begin{equation}
H_{\text{g}}=\sum_{s\sigma}\int^{k_{c}}_{-k_{c}}dk\varepsilon_{k}c^{\dagger}_{sk\sigma}c_{sk\sigma},
\label{hg5}
\end{equation}
where $\varepsilon_{k}=\hbar v_{F}k$, $s$ stands for the valley index, $k_{c}$ is a cut-off of the momentum so that $D=\hbar v_{F}k_{c}$ where $D$ is the cut-off of the energy (the explicit value of $D$ will be given below). The Hamiltonian of the impurity is taken as  $H_{\text{im}}=\varepsilon_{0}\sum_{\sigma}f^{\dagger}_{\sigma}f_{\sigma}+\frac{U}{2}\sum_{\sigma}n_{\sigma}n_{\bar{\sigma}}$, where $f_{\sigma}$ is the annihilation operator; here the spin $\sigma$ is at the impurity site, and $n_{\sigma}=f^{\dagger}_{\sigma}f_{\sigma}$. The total Hamiltonian is $H=H_{\text{g}}+H_{\text{im}}+H_{\text{c}}$, where the hybridization Hamiltonian is
\begin{equation}
H_{\text{c}}=\tilde{v}_{0}\sum_{s\sigma}\int^{k_{c}}_{-k_{c}}dk\sqrt{|k|}\left(c^{\dagger}_{sk\sigma}f_{\sigma}+f^{\dagger}_{\sigma}c_{sk\sigma}\right),
\label{hc2}
\end{equation}
with $\tilde{v}_{0}=\frac{v_{0}\sqrt{\pi\Omega_{0}}}{2\pi}$, $\Omega_{0}$ is the unit cell area, and $v_{0}$ is the hybridization strength in energy dimension.

\subsection{The Green's function and the selfenergies}
\subsubsection{The first order approach}
We define the retarded Green's function $\langle\langle A(t)|B(0)\rangle\rangle^{r}=-i\theta(t)\langle\{A(t),B(0)\}\rangle$ \cite{zubarev,gf}, where $A(B)$ are operators. $\langle\langle A|B\rangle\rangle_{\omega^{+}}$ is the  ${\omega^{+}}$ transform  of $\langle\langle A(t)|B(0)\rangle\rangle^{r}$, where $\omega^{\pm}=\omega\pm i0^{+}$. In the following, we calculate the retarded Green's function and symbol the index  $\omega^{+}$ as $\omega$ so that we can obtain the retarded (advanced) Green's function when replacing $\omega$ by $\omega^{\pm}$ in the derived Green's function. The standard equation of motion \cite{zubarev,gf} of the Green's function reads
$\omega\langle\langle A|B\rangle\rangle_{\omega}=\langle\{A,B\}\rangle+\langle\langle[A,H]|B\rangle\rangle_{\omega}$. We  then calculate the Green's function for the impurity $t_{\sigma\sigma'}(\omega)=\langle\langle f_{\sigma}|f_{\sigma'}^{\dagger}\rangle\rangle_{\omega}$ by using
\begin{equation}
\omega\langle\langle
f_{\sigma}|f_{\sigma'}^{\dagger}\rangle\rangle_{\omega}=\langle\{f_{\sigma},f_{\sigma'}^{\dagger}\}\rangle+\langle\langle[f_{\sigma},H]|f_{\sigma'}^{\dagger}\rangle\rangle_{\omega},
\label{gfff}
\end{equation}
where
\begin{equation}
[f_{\sigma},H]=\epsilon_{0}f_{\sigma}+Uf_{\sigma}n_{\bar{\sigma}}+\tilde{v}_{0}\sum_{s'}\int dk'\sqrt{|k'|}c_{s'k'\sigma}.
\label{fh}
\end{equation}
Thus, we find
\begin{eqnarray}
(\omega-\epsilon_{0})t_{\sigma\sigma'}(\omega) &=& \delta_{\sigma\sigma'}+U\langle\langle
f_{\sigma}n_{\bar{\sigma}}|f_{\sigma'}^{\dagger}\rangle\rangle_{\omega} \notag\\
&+&\tilde{v}_{0}\sum_{s}\int dk\sqrt{|k|}\langle\langle
c_{sk\sigma}|f_{\sigma'}^{\dagger}\rangle\rangle_{\omega}. 
\label{gfff1}
\end{eqnarray}
%
%

\begin{figure}[tbh]
\includegraphics[width=0.35\textwidth]{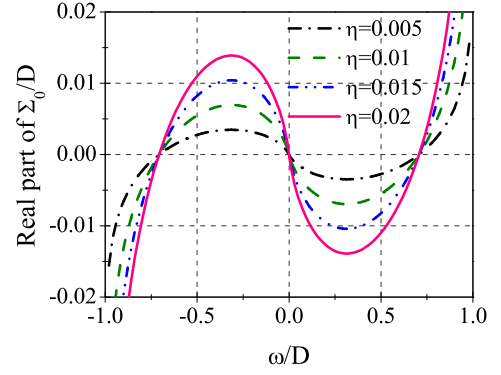}
\caption{(color online) The real part of $\Sigma_{0}(\omega^{+})$, as  given by Eq. (\ref{selfenergy1}).  \label{realpart}}
\end{figure}

By employing the equation
\begin{equation}
[c_{sk\sigma},H]=\varepsilon_{k}c_{sk\sigma}+\tilde{v}_{0}\sqrt{|k|}f_{\sigma},
\label{ch}
\end{equation}
we obtain
\begin{equation}
\langle\langle
c_{sk\sigma}|f_{\sigma'}^{\dagger}\rangle\rangle_{\omega}=\frac{\tilde{v}_{0}\sqrt{|k|}}{\omega-\varepsilon_{k}}t_{\sigma\sigma'}.
\label{cf}
\end{equation}
Therefore, the third term on the right hand side of Eq. (\ref{gfff1}) reads $\Sigma_{0}(\omega)t_{\sigma\sigma'}(\omega)$, where the selfenergy is defined by $\Sigma_{0}(\omega^{+})=\tilde{v}^{2}_{0}\sum_{s}\int dk|k|(\omega^{+}-\varepsilon_{k})^{-1}$ and is readily calculated as \cite{zhu2011}
\begin{equation}
\Sigma_{0}(\omega^{+})=-\eta\left[\omega\ln\frac{|D^{2}-\omega^{2}|}{\omega^{2}}+i\pi|\omega|\theta(D-|\omega|)\right], \label{selfenergy1}
\end{equation}
where 
$\eta=\frac{\Omega_{0}}{2\pi}\frac{v^{2}_{0}}{(\hbar v_{F})^{2}}$. Let us discuss the structure of
 this selfenergy  (see Fig. \ref{realpart}). The imaginary part is simply proportional to the absolute value of the energy with respect to the Dirac point, i.e. $|\omega|$. The real part is shown in Fig. \ref{realpart}. When $\omega\rightarrow0$, the real part tends to zero as well which is a typical character in the marginal Fermi liquid theory \cite{zhanggm}. When $\omega$  approaches the band edges, a singularity occurs. A similar singularity at the  band edge is met in the impurity problem for a normal metal \cite{hewson}. The difference is that the parity in energy space (with respect to the Fermi level that is  assumed to be in the middle of the band) is even (odd) in a normal metal (graphene). Another observation is that the zero points occur at $\omega=0$ and $\omega=\pm\frac{\sqrt{2}}{2}D$.

Having obtained the selfenergy we can write down the equation
\begin{equation}
[\omega-\varepsilon_{0}-\Sigma_{0}(\omega)]t_{\sigma\sigma'}=\delta_{\sigma\sigma'}+UG(\omega),
\label{tg}
\end{equation}
where $G(\omega)=\langle\langle f_{\sigma}n_{\bar{\sigma}}|f_{\sigma'}^{\dagger}\rangle\rangle_{\omega}$.
%
We can further formulate the equation of $G$ as
\begin{eqnarray}
(\omega-\varepsilon_{1})G(\omega) &=& \langle n_{\bar{\sigma}}\rangle\delta_{\sigma\sigma'}+\tilde{v}_{0}\sum_{s}\int dk\sqrt{|k|}[F_{1}(\omega) \notag\\
&+& F_{2}(\omega)-F_{3}(\omega)],
\label{gf123}
\end{eqnarray}
where $\varepsilon_{1}=\varepsilon_{0}+U$, $F_{1}(\omega)=\langle\langle n_{\bar{\sigma}}c_{sk\sigma}|f^{\dagger}_{\sigma'}\rangle\rangle_{\omega}$, $F_{2}(\omega)=\langle\langle f^{\dagger}_{\bar{\sigma}} c_{sk\bar{\sigma}}f_{\sigma}|f^{\dagger}_{\sigma'}\rangle\rangle_{\omega}$, and $F_{3}(\omega)=\langle\langle c^{\dagger}_{sk\bar{\sigma}}f_{\bar{\sigma}} f_{\sigma}|f^{\dagger}_{\sigma'}\rangle\rangle_{\omega}$ are higher order Green's functions.

\subsubsection{Accounting for higher order terms}
It is straightforward to obtain the following equations in the same way for the higher order Green's functions
\begin{eqnarray}
&&(\omega-\varepsilon_{k})F_{1}=\tilde{v}_{0}\sqrt{|k|}G+\tilde{v}_{0}\sum_{s'}\int dk'\sqrt{|k'|}\times \notag\\
&&\left(\langle\langle f^{\dagger}_{\bar{\sigma}}c_{s'k'\bar{\sigma}}c_{sk\sigma}|f^{\dagger}_{\sigma'}\rangle\rangle-\langle\langle c^{\dagger}_{s'k'\bar{\sigma}}f_{\bar{\sigma}}c_{sk\sigma}|f^{\dagger}_{\sigma'}\rangle\rangle\right),
\label{f1}
\end{eqnarray}
\begin{eqnarray}
(\omega-\varepsilon_{k})F_{2} &=& \tilde{v}_{0}\sqrt{|k|}G+\langle f^{\dagger}_{\bar{\sigma}}c_{sk\bar{\sigma}}\rangle\delta_{\sigma\sigma'} \notag \\
&+& \tilde{v}_{0}\sum_{s'}\int dk'\sqrt{|k'|}\left(\langle\langle f^{\dagger}_{\bar{\sigma}}c_{sk\bar{\sigma}}c_{s'k'\sigma}|f^{\dagger}_{\sigma'}\rangle\rangle\right. \notag\\ 
&-&\left.\langle\langle c^{\dagger}_{s'k'\bar{\sigma}}c_{sk\bar{\sigma}}f_{\sigma}|f^{\dagger}_{\sigma'}\rangle\rangle\right),
\label{f2}
\end{eqnarray}
and
\begin{eqnarray}
&&(\omega-\varepsilon_{2}+\varepsilon_{k})F_{3} = -\tilde{v}_{0}\sqrt{|k|}G+\langle c^{\dagger}_{sk\bar{\sigma}}f_{\bar{\sigma}}\rangle\delta_{\sigma\sigma'} \notag \\
&&+\tilde{v}_{0}\sum_{s'}\int dk'\sqrt{|k'|}\left(\langle\langle c^{\dagger}_{sk\bar{\sigma}}c_{s'k'\bar{\sigma}}f_{\sigma}|f^{\dagger}_{\sigma'}\rangle\rangle\right. \notag\\
&&+\left.\langle\langle c^{\dagger}_{sk\bar{\sigma}}f_{\bar{\sigma}}c_{s'k'\sigma}|f^{\dagger}_{\sigma'}\rangle\rangle\right),
\label{f3}
\end{eqnarray}
where $\varepsilon_{2}=2\varepsilon_{0}+U$ is the energy for the double occupation of the impurity.
As usual, to obtain useful expressions we have to truncate the iterative equation by some approximations. We follow the
standard method in Refs. \cite{lacroix,theumann,appelbaum} by
assuming that the operator pair with the same spin indices can be pulled
out of the Green's function (GF) and regarded as an average. This
assumption amounts to considering only spin conserved
couplings within graphene or between graphene and the impurity. Physically, this amounts  to saying  that there is no  mechanism to
flip the spin when it hops from the impurity to graphene or when it propagates  in
graphene. The equation of motion method is able to capture the Kondo effect only qualitatively but can unravel explicitly how the Kondo resonance develops in the local density of state \cite{gf,eom}.
 We show one example of the decoupling scheme in the equation of $F_{1}$, i.e.
\begin{equation}
\langle\langle f^{\dagger}_{\bar{\sigma}}c_{s'k'\bar{\sigma}}c_{sk\sigma}|f^{\dagger}_{\sigma'}\rangle\rangle_{\omega}\approx\langle f^{\dagger}_{\bar{\sigma}}c_{s'k'\bar{\sigma}}\rangle\langle\langle c_{sk\sigma}|f^{\dagger}_{\sigma'}\rangle\rangle_{\omega}.
\label{approximation}
\end{equation}
The others can be done in the same way. After straightforward manipulations, we can find the GF
\begin{equation}
t_{\sigma\sigma'}=\delta_{\sigma\sigma'}\frac{1+(U/\tilde{\omega})(\langle n_{\bar{\sigma}}\rangle+A_{\bar{\sigma}})}{\omega-\varepsilon_{0}-\Sigma_{0}-(U/\tilde{\omega})(\Sigma_{0}A_{\bar{\sigma}}-\Sigma_{1\bar{\sigma}})},
\label{tgf}
\end{equation}
where $\tilde{\omega}=\omega-\varepsilon_{1}-\Sigma_{2}$, $\Sigma_{2}=2\Sigma_{0}(\omega)-\Sigma_{0}(\varepsilon_{2}-\omega)$,
\begin{equation}
A_{\bar{\sigma}}(\omega)=\tilde{v}_{0}\sum_{s}\int dk\sqrt{|k|}\left(\frac{\langle f^{\dagger}_{\bar{\sigma}}c_{sk\bar{\sigma}}\rangle}{\omega-\varepsilon_{k}}-\frac{\langle c^{\dagger}_{sk\bar{\sigma}}f_{\bar{\sigma}}\rangle}{\omega-\varepsilon_{2}+\varepsilon_{k}}\right),
\label{afunction}
\end{equation}
and
\begin{eqnarray}
\Sigma_{1\bar{\sigma}}(\omega) &=& \tilde{v}^{2}_{0}\sum_{ss'}\int dkdk'\sqrt{|k|}\sqrt{|k'|}\left(\frac{\langle c^{\dagger}_{s'k'\bar{\sigma}}c_{sk\bar{\sigma}}\rangle}{\omega-\varepsilon_{k}}\right. \notag\\
&+& \left.\frac{\langle c^{\dagger}_{sk\bar{\sigma}}c_{s'k'\bar{\sigma}}\rangle}{\omega-\varepsilon_{2}+\varepsilon_{k}}\right).
\label{self1}
\end{eqnarray}

We employ the relation between the average and the GF
\begin{equation}
\langle ab\rangle=-\frac{1}{\pi}\Im\int d\omega'f(\omega')\langle\langle
b|a\rangle\rangle_{\omega'^{+}}, \label{gfaverage}
\end{equation}
where $f(\omega')$ is the Fermi function. Here we introduced $\omega'^{\pm}=\omega'\pm i0^{+}$ for the retarded (advanced) GF. It is easy to show
$\langle f^{\dagger}_{\bar{\sigma}}c_{sk\bar{\sigma}}\rangle=\langle c^{\dagger}_{sk\bar{\sigma}}f_{\bar{\sigma}}\rangle$. And the GF of the graphene with an impurity has the form
\begin{eqnarray}
\langle\langle c_{sk\sigma}|c^{\dagger}_{s'k'\sigma'}\rangle\rangle_{\omega} &=& \frac{\delta_{\sigma\sigma'}\delta_{ss'}\delta(k-k')}{\omega-\varepsilon_{k}} \notag\\
&+& t_{\sigma\sigma'}\frac{\tilde{v}^{2}_{0}\sqrt{|k||k'|}}{(\omega-\varepsilon_{k})(\omega-\varepsilon_{k'})}.
\label{ccfunction}
\end{eqnarray}
In the calculation of the selfenergy $\Sigma_{1\bar{\sigma}}$ we only keep the first term in Eq. (\ref{ccfunction}), i.e. we use the GF of the pure graphene system without the impurity. In the following, we only focus on the limit $U\rightarrow\infty$ in which case we infer the expression
\begin{equation}
t_{\sigma\sigma'}=\delta_{\sigma\sigma'}\frac{1-\langle n_{\bar{\sigma}}\rangle-A_{\bar{\sigma}}}{\omega-\varepsilon_{0}-(1-A_{\bar{\sigma}})\Sigma_{0}-\Sigma_{1\bar{\sigma}}}.
\label{tgf1}
\end{equation}
The selfenergy $\Sigma_{1\bar{\sigma}}$ can be calculated exactly (see Appendix)
\begin{eqnarray}
\Sigma_{1\bar{\sigma}}(\omega^{+}) &=& \eta\left[\frac{|D|}{2}\ln\frac{|\omega^{2}-D^{2}|}{(2\pi T)^{2}}-|\mu|\psi\left(\frac{1}{2}+\frac{\omega-\mu}{2\pi iT}\right)\right] \notag\\
&+& \frac{1}{2}\Sigma_{0}(\omega^{+}), \label{self12}
\end{eqnarray}
where $\psi$ is the  Digamma function.
\subsubsection{Calculations of the spectral function $A_{\bar{\sigma}}(\omega)$}
An evaluation of the function $A_{\bar{\sigma}}(\omega)$ yields
\begin{eqnarray}
A_{\bar{\sigma}}(\omega) &=& -\frac{1}{2\pi}\int d\omega'f(\omega')\tilde{v}_{0}^{2}\sum_{s}\int dk|k| \notag\\
&\times &\left[\frac{t_{\bar{\sigma}\bar{\sigma}}(\omega'^{+})}{(\omega-\varepsilon_{k}+i0^{+})(\omega'-\varepsilon_{k}+i0^{+})}\right. \notag\\
&-&\left.\frac{t^{*}_{\bar{\sigma}\bar{\sigma}}(\omega'^{+})}{(\omega-\varepsilon_{k}+i0^{+})(\omega'-\varepsilon_{k}-i0^{+})}\right].
\label{afunction1}
\end{eqnarray}
The denominator in the first term can be split into
\begin{equation}
\frac{1}{\omega'-\omega}\left[\frac{1}{\omega-\varepsilon_{k}+i0^{+}}-\frac{1}{\omega'-\varepsilon_{k}+i0^{+}}\right].
\notag
\end{equation}
Because of the term  $\omega'-\omega$ in the denominator
the integral is peaked at  $\omega'\approx\omega$.
However, the difference of the two terms in the square brackets will disappear in this case. Therefore, the whole contribution of this integral is less important than the second term in Eq. (\ref{afunction1}) and can be neglected. Splitting the product in the denominator of the second term in Eq. (\ref{afunction1}), we find
\begin{equation}
A_{\bar{\sigma}}(\omega^{+})=(2\pi)^{-1}\int\frac{d\omega'f(\omega')t^{*}_{\bar{\sigma}\bar{\sigma}}(\omega'^{+})
[\Sigma_{0}(\omega^{+})-\Sigma_{0}(\omega'^{-})]}
{\omega'-\omega-i0^{+}}.
\label{afunction2}
\end{equation}
In accordance with the consideration that the most contribution comes from $\omega'\approx\omega$, we may simplify the expression
 further by using
\begin{equation}
\Sigma_{0}(\omega^{+})-\Sigma_{0}(\omega'^{-})\approx -2\pi\eta|\omega'|\theta(D-|\omega'|),
\label{selfdiffer}
\end{equation}
to obtain
\begin{equation}
A_{\bar{\sigma}}(\omega^{+})=-\eta\int_{-D}^{D}\frac{d\omega'|\omega'|f(\omega')t^{*}_{\bar{\sigma}\bar{\sigma}}(\omega'^{+})}{\omega'-\omega-i0^{+}}.
\label{afunction3}
\end{equation}
A similar relation was derived for a normal metal as a host material in Ref. \cite{lacroix}.

At low temperatures, by using the same assumption as in Ref. \cite{lacroix}, we treat the GF as being a smoother part than the other term in the integrand. Thus, GF can be pulled out of the integral and it depends on $\omega$ rather than on $\omega'$. Doing so yields
\begin{equation}
A_{\bar{\sigma}}(\omega^{+})=t^{*}_{\bar{\sigma}\bar{\sigma}}(\omega^{+})\Sigma_{1\bar{\sigma}}(\omega^{+}).
\label{afunction4}
\end{equation}

At high temperatures, the function $A$ tends to be small. We assume a zero order Green's function, i.e. $t_{\bar{\sigma}\bar{\sigma}}(\omega'^{+})=(1-\langle n_{\sigma}\rangle)/(\omega'^{+}-\varepsilon_{0})$, and substitute it into Eq. (\ref{afunction3}) leading to the
 compact formula of the
function $A$ at high temperatures
\begin{equation}
A_{\bar{\sigma}}(\omega^{+})=-\frac{(1-\langle n_{\sigma}\rangle)[\Sigma_{1\bar{\sigma}}(\omega^{+})-\Sigma_{1\bar{\sigma}}(\varepsilon_{0}^{+})]}{\varepsilon_{0}-\omega}.
\label{afunction5}
\end{equation}
\section{Discussions and Numerical calculations}
\subsection{The case of a vanishing $A$-function}
\subsubsection{The Kondo resonance and the Kondo temperature}
From the calculation of the function $A$ we concluded that it is related to the selfenergy $\Sigma_{1\bar{\sigma}}$. It is necessary to investigate the singularity of the $A$-function  but we postpone it to the next subsection. From Eq. (\ref{afunction4}), we know that the only  possibility
of having a  singularity in $A$ is via the GF $t_{\bar{\sigma}\bar{\sigma}}$ since the selfenergy $\Sigma_{1\bar{\sigma}}$ is non-singular,  except for $T\rightarrow 0$ (and we do not consider the zero temperature).
 For a comparison, the corresponding  selfenergy $\Sigma_{1\bar{\sigma}}$ for a normal metal
has a singularity when $\omega\rightarrow\mu$ at low temperatures.

\begin{figure}[tbh]
\includegraphics[width=0.44\textwidth]{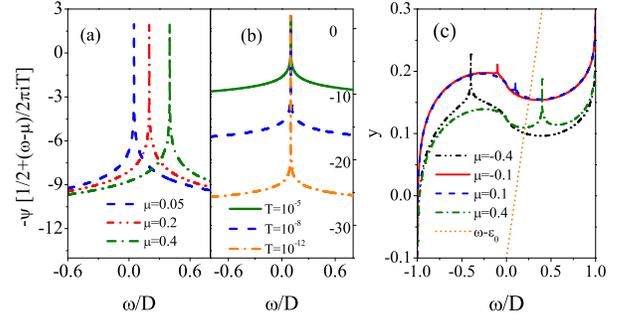}
\caption{(color online) The digamma function is shown with different gate voltages and temperatures in (a) and (b), respectively. In (a), $T=10^{-5} D$, and $\eta=0.02$. In (b), $\mu=0.1 D$ and $\eta=0.02$.
A graphic method to solve  Eq. (\ref{condition1}) is shown in (c). The short-dash straight line is for the left hand side of Eq. (\ref{condition1}) where $\varepsilon_{0}=0.1$. The curves are for the right hand side of Eq. (\ref{condition1}) where the digamma function is calculated numerically. The other parameters are $\eta=0.02$ and $T=10^{-5} D$. The energy unit is taken as $D$ in all graphs.
\label{digammafig}}
\end{figure}
We note that the function  $A$ depends linearly on  $\eta$ which is a small quantity for the graphene case. If the function $A$ is not singular exactly,  we can neglect it in the numerator and denominator since the terms with $A$ are at least one order of magnitude  less than the other terms (in $\eta$). This approximation yields a zero order result and we infer, in the following, that it does not change  qualitatively the buildup of the Kondo resonance. Thus, here we study the case of a vanishing   $A$-function. The Green's function now reads
\begin{eqnarray}
&& t_{\sigma\sigma'}= \label{tgf}\\
&&\frac{\delta_{\sigma\sigma'}(1-\langle n_{\bar{\sigma}}\rangle)}{\omega-\varepsilon_{0}-1.5\Sigma_{0}-\eta\left[\frac{|D|}{2}\ln\frac{|\omega^{2}-D^{2}|}{(2\pi T)^{2}}-|\mu|\psi\left(\frac{1}{2}+\frac{\omega-\mu}{2\pi iT}\right)\right]}. \notag
\end{eqnarray}
The resonances are determined by the positions where the real part vanishes, i.e.
\begin{eqnarray}
\omega-\varepsilon_{0} &=& 1.5\Re\Sigma_{0}+\eta\left[\frac{|D|}{2}\ln\frac{|\omega^{2}-D^{2}|}{(2\pi T)^{2}}\right. \notag\\
&-& \left.|\mu|\Re\psi\left(\frac{1}{2}+\frac{\omega-\mu}{2\pi iT}\right)\right].
\label{condition1}
\end{eqnarray}

The Kondo resonance manifests itself in the solution around the Fermi level and is largely determined by the behavior of the digamma function. We thus study the behavior of digamma function at first. In Fig. \ref{digammafig}(a) and (b), the digamma function variation with $\omega$ is shown for different gate voltages and temperatures. Compared with the results for a normal metal, we observe  new features.\\
 i) The $\eta$ is small in graphene while it is larger in normal metal because of the constant large DOS. ii) A factor $|\mu|$ appears in front of the digamma function which suppresses the resonant effect when $\mu$ tends to half-filling ($\mu=0$). This means the Kondo resonance is suppressed when the gate voltage is small and eventually disappears at half-filling. This linear dependence on $|\mu|$ is just a consequence of the relativistic dispersion of graphene. A another pronounced effect of it is  that the temperature dependence in the second term and the third term of the right hand side of the Eq. (\ref{condition1}) do not cancel with each other. The consequence is that there is no  singularity when $\omega$ tends to the Fermi level. This is simply understood. For a very small but finite temperature, if $\omega\rightarrow\mu$ then  $\psi\rightarrow\psi(1/2)$ rather than a logarithmic divergence in normal metal case (see Ref. \cite{lacroix}). This is reflected by the uniform topmost value of the peaks in Fig. \ref{digammafig}(a) and (b), i.e. $-\psi(1/2)\approx1.96351$. The peaks are emerging at the Fermi level which can be tuned with the gate voltage (see Fig. \ref{digammafig}(a)). The peaks only exist in a narrow range of $\omega$ in which $|\omega-\mu|$ can be comparable to $2\pi T$. It decreases dramatically outside this interval. When $|\omega-\mu|\gg 2\pi T$, the real part of the digamma function can be approximated by a logarithmic function, i.e. $\Re\psi=\ln\frac{\sqrt{(\omega-\mu)^{2}+(\pi T)^{2}}}{2\pi T}$. In this case, the digamma function develops a singularity  only when $T\rightarrow 0$. Please, note that this singularity in temperature is removed by the second part of Eq. (\ref{condition1}) for a normal metal. However, the gate-voltage dependence of the digamma function lets this term survive (see Fig. \ref{digammafig}(b)).

Because of the smallness of $\eta$ and the non-singular peak located at the Fermi level, it is relevant to inspect whether a solution around the Fermi level (i.e. the Kondo resonance) can be obtained. For a gate voltage $|\mu|<D$, $\Re\Sigma_{0}$ does not give rise to a singularity. The second term in Eq. (\ref{condition1}), with the logarithmic behavior increases with lowering the temperature rendering possible a Kondo resonance. To show this clearly, we resort to a graphic method to solve Eq. (\ref{condition1}) in Fig. \ref{digammafig}(c). Whenever an intersection of the peaks of the curves and the straight line occurs, the Kondo resonance emerges. We should emphasize that the vanishing DOS of graphene at the Dirac points at half-filling suppresses the Kondo resonance. To release this suppression, a reasonable gate voltage is necessary. As a consequence of the non-singular peak at the  Fermi level, we infer the important conclusion that the Kondo resonance can only be observed in  a quite narrow window of $\varepsilon_{0}$ or the gate voltage with respect to the Fermi energy (at a fixed low temperature).

\begin{figure}[tbh]
\includegraphics[width=0.4\textwidth]{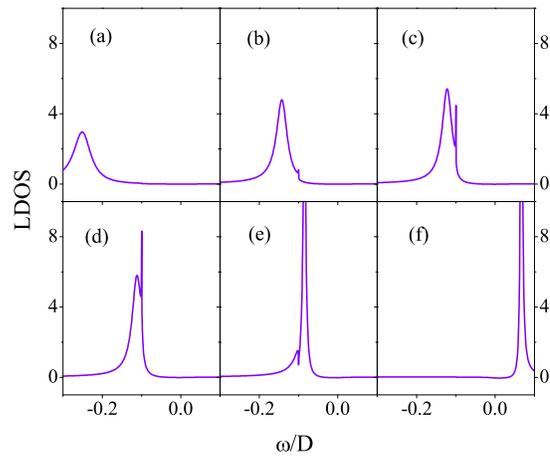}
\caption{(color online) LDOS, evaluated as $D\mathcal{N}_{f\sigma}$, is shown for different $\varepsilon_{0}$ in (a) through (f). $\mu=-0.1$, $T=10^{-5}$, and $\eta=0.02$. $\varepsilon_{0}=-0.45,-0.34,-0.32,-0.31,-0.28$, and $-0.1$ in (a) to (f) respectively. Energy is measured in $D$. $n_{\sigma}$ is assumed to be 0.5. \label{ldosfig}}
\end{figure}
\textit{Kondo temperature}: If the temperature is very low and $\beta(\omega-\mu)\gg1$, we may obtain a virtual level as
\begin{equation}
\varepsilon'_{0}=\varepsilon_{0}+1.5\Re\Sigma_{0}+\eta|\mu|\left(\ln\frac{D^{D/|\mu|}(2\pi T)^{1-D/|\mu|}}{|\varepsilon'_{0}-\mu|}\right).
\label{virtuallevel}
\end{equation}
Since we seek a  solution for the Kondo resonance near the Fermi level, we may from Eq. (\ref{virtuallevel}) obtain that
\begin{equation}
|\varepsilon'_{0}-\mu|\approx D^{D/|\mu|}(2\pi T)^{1-D/|\mu|}e^{\frac{\varepsilon_{0}+1.5\Re\Sigma_{0}-\mu}{\eta|\mu|}}.
\label{virtual1}
\end{equation}
The estimation of the Kondo temperature in this case should be supplemented by studying  the scaling process \cite{haldane}.
In the scaling method, the bandwidth $D$ is reduced to an effective bandwidth $\tilde{D}$ and the Hamiltonian is renormalized with a new virtual level. This reduction can be done until  $\tilde{D}$ reaches the Fermi energy. In view of this picture, we take $D\rightarrow\tilde{D}\sim|\mu|$ and obtain the Kondo temperature $T_{\text{K}}$
\begin{equation}
|\varepsilon'_{0}-\mu|\approx|\mu|e^{\frac{\varepsilon_{0}-\mu}{\eta|\mu|}}\approx T_{\text{K}}.
\label{kondotem}
\end{equation}
 The real part of $\Sigma_{0}$ is ignored for the case where $|\mu|$ is not so large. There is no difficulty to include the $\Sigma_{0}$ by setting its argument $\omega=\mu$. When $|\mu|$ tends to zero, the Kondo temperature tends to zero for $\varepsilon_{0}<\mu$. This is a natural observation from Fig. \ref{digammafig}(c), and can be understood by the fact that no solution around the Fermi level can be obtained for a vanishing digamma function in this case. The intuitive picture is, there are no free electrons to screen the spin,  since the local density of state is zero. With increasing $|\mu|$, the Kondo temperature increases exponentially \cite{zhu10} with a linear prefactor in $|\mu|$. This build up  is qualitatively consistent with the prediction made on the basis of the  \textit{sd} model \cite{sengupta}.

In Ref. \cite{vojta}, the authors considered   2D Dirac fermions with the linear DOS $|\omega|^{r}$ where $r$ is a dimension index ($r=1$ for graphene). They studied the scaling and revealed that $T_{+\text{K}}(r\rightarrow 1)=|\mu|e^{-1.7/(1-r)}$ and $T_{-\text{K}}(r\rightarrow 1)=|\mu|[W(1+r)/(1+r)]^{1/r}$, where $\pm$ for $\mu\gtrless0$. These results also show that the Kondo effect is suppressed when $\mu\rightarrow 0$. 
By quoting the data from Ref. \cite{vojta},
i.e. $T_{\text{K}}\approx15 K$ at $|\mu|=0.2$ eV, we estimate the impurity level to be $\varepsilon_{0}\approx\mu-5\eta|\mu|$. We note $\eta|\mu|$ characterizes the hybridization strength at the Fermi level $\mu$. We first wish  to compare this coupling with that characteristic for a normal metal or  semiconductor quantum dot systems, i.e. $\Delta=\pi v^{2}_{0}/(2D)$ \cite{lacroix}. The ratio is $\frac{\eta|\mu|}{2\Delta}=\frac{D|\mu|}{(13.6 eV)^{2}}$, where $a=1.42 {\AA}$ and $v_{F}=10^{6}$ m/s have been used \cite{graphene}. If we take $D\approx 7$ eV \cite{uchoa11} and $|\mu|=0.2$ eV, the ratio is about 0.0035. Therefore, we infer that the coupling between graphene and the adatom impurity is  weaker than  in the case of a normal metal host. Secondly, we estimate the value of $\eta$. If $v_{0}=1$ eV \cite{uchoa11}, we find  $\eta\approx 0.0085$. Thus, we estimate for the impurity level $\varepsilon_{0}=\mu-8.5$ meV (for $|\mu|=0.2$ eV). The impurity level is hence quite close to the Fermi energy for an observation of the Kondo resonance.


\subsection{LDOS of the impurity } 

Let us  define the local density of state (LDOS) for the spin $\sigma$ as $\mathcal{N}_{f\sigma}(\omega)=-\frac{\Im}{2\pi}t_{\sigma\sigma}(\omega^{+})$,
 and show how the Kondo resonance develops and disappears when varying the impurity level in Fig. \ref{ldosfig}.
We note that ignoring the function $A$ in the numerator could induce a quantitative difference in the calculated occupation, even when  $A$ is small. However, in this case it will not induce a qualitative shift since it does not contribute significantly  to the  Kondo resonance. In the denominator, the term with the function  $A$ and $\Sigma_{0}$ can be neglected since this is a second order term
  in $\eta$, while $\eta$ is much smaller than its counterpart in a normal metal. Under these considerations, we performed numerical calculations of the LDOS; the results are shown in Fig. \ref{ldosfig}.

 In Fig. \ref{ldosfig}(a) through (f), the impurity level energy is  increased gradually.
 In (a), the Kondo resonance is not yet developed. In (b) and (c), it starts emerging, and in (d) a pronounced Kondo peak has developed.
  In (e), when the virtual main peak crosses  the Kondo resonance, the mutual interplay shows as a dip. In (f), the virtual main peak passes over the Fermi level. As a consequence, the Kondo resonance disappears. This build-up of the Kondo resonance shown in Fig. \ref{ldosfig} is also valid if we view the increase of the impurity level energy  as a lowering the Fermi level (while keeping fixed the impurity level energy). Therefore, our findings  can be tested easily by tuning the gate voltage in the experiments.

To gain more insights in the specific character of the Kondo resonance in graphene, we compare what we observe here with the normal metal case where the Kondo temperature is $T_{\text{K}}\sim(D\Delta)^{1/2}\exp[\frac{\pi(\varepsilon_{0}-\mu)}{2\Delta}]$ in the $U\rightarrow\infty$ limit \cite{hewson,lacroix}. When $\mu$ is fixed, the Kondo temperature is decreasing when moving  $\varepsilon_{0}$ away from $\mu$ (below it). The Kondo resonance in  LDOS  disappears when $T_{\text{K}}$ is below $T$, similar to the behaviour observed  in Fig. \ref{ldosfig}. The difference  to the normal metal case is in the  parameter window for $\varepsilon_{0}$ where  Kondo resonance exists. This
window is for graphene  rather narrow compared with the normal metal, as the ratio   $\frac{\eta|\mu|}{2\Delta}$ is quite small (about 0.0035, as given above).  According to the parameters used in  Fig. \ref{ldosfig} and to keep $T\leqslant T_{\text{K}}$, the parameter window for $\varepsilon_{0}$ derived by using  Eq. (\ref{kondotem}) should be in a range of  $9\eta|\mu|\approx0.02$ below the Fermi level which is consistent with the observation from the figure.
   %


\begin{figure}
\includegraphics[width=0.4\textwidth]{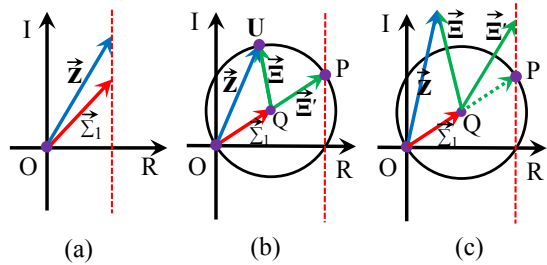}
\caption{(color online) A geometrical analysis for the Kondo resonance and the singularity of the Green function is shown. In (a) the function $A$ vanishes. In (b) and (c)  $A$ is finite. The dashed vertical lines indicate the positions where the resonances can occur. R (I) on the axis means real (imaginary) axis. In (b) and (c), the rings are the singular rings with poles of the Green function.
\label{circle}}
\end{figure}

\subsection{Non-zero $A$ function study}
In this subsection, we wish to discuss the situation where the function $A$ is solved by means of GF itself, i.e.  Eq. (\ref{afunction4}). Thus, we find an iterative equation for $t_{\bar{\sigma}\bar{\sigma}}$ and the $A_{\bar{\sigma}}$-function. We consider the spin degenerate case and infer
\begin{equation}
t=\frac{m-t^{*}\Sigma_{1}}{\Xi+t^{*}\Sigma_{0}\Sigma_{1}},
\label{gfanzero0}
\end{equation}
where $\Xi=z-\Sigma_{1}$, $z=\omega-\varepsilon_{0}-\Sigma_{0}$, $m=1-\langle n\rangle$ and $n=n_{\uparrow}=n_{\downarrow}$. By substituting $t^{*}$, we obtain the GF
\begin{eqnarray}
t &=& \frac{-\Lambda\pm\sqrt{\Gamma}}{2(\Xi\Sigma^{*}_{0}-\Sigma_{0}\Sigma_{1})\Sigma^{*}_{1}}, \notag\\
\Lambda &=& (|\Xi|^{2}-|\Sigma_{1}|^{2})+2im\Im(\Sigma_{0}\Sigma_{1}), \notag\\
\Gamma &=& (|\Xi|^{2}-|\Sigma_{1}|^{2})^{2}+4m(|\Xi|^{2}+|\Sigma_{1}|^{2})\Re(\Sigma_{0}\Sigma_{1}) \notag\\
&-&4m^{2}[\Im(\Sigma_{0}\Sigma_{1})]^{2}-8m|\Sigma_{1}|^{2}\Re(\Xi\Sigma^{*}_{0}).
\label{gfanzero1}
\end{eqnarray}
Please, note that $\Gamma$ is a real function. The singularity occurs when
\begin{equation}
\Xi\Sigma^{*}_{0}-\Sigma_{0}\Sigma_{1}=0.
\label{singularcondition}
\end{equation}
By using the conjugate relation $\Xi^{*}\Sigma_{0}=\Sigma^{*}_{0}\Sigma^{*}_{1}$ we obtain the constraint
\begin{equation}
|\Xi|=|\Sigma_{1}|,
\label{singularcondition1}
\end{equation}
that leads to a singularity. To discuss this equation, let us first inspect the singularity analysis for a vanishing  $A$.
To do so we view a complex function as a vector in the complex plane (cf. Fig. \ref{circle}). The singularity in this case is given by $|\vec{\Xi}|=0$ or $\vec{z}=\vec{\Sigma}_{1}$. That means, the vector $\vec{z}$ coincides with $\vec{\Sigma}_{1}$. However, when they are not completely coincident
  but merely  have the same real part (which is shown by the two vectors reaching  the vertical dashed line in Fig. \ref{circle}(a))
  the vanishing real part gives rise to a prominent resonance.
   If it happens at the Fermi level, we observe a Kondo resonance.

For nonzero $A$, by noting the fact that $\Xi=z-\Sigma_{1}$, the relation in Eq. (\ref{singularcondition1}) reflects a geometrical relation shown in Fig. \ref{circle}(b). A singular ring is generated with the radius set by  the vector $\vec{\Sigma}_{1}$ and we have $\vec{\Xi}=\vec{z}-\vec{\Sigma}_{1}$. The basic physics in this singular ring is that the vector $\vec{z}$ and $\vec{\Sigma}_{1}$ are varying with $\omega$. At some particular values, the endpoint of the vector $\vec{z}$ lies  on the ring defined by  $|\vec{\Sigma}_{1}|$. A singularity is then encountered. Thus, the ring reflects the singularity of the GF in our investigation.
Considering the point "O" on the ring, it means $\vec{\Xi}=-\vec{\Sigma}_{1}$ or $\vec{z}=0$. Thus it reflects a resonance developed at the main peak of LDOS sitting at a renormalized virtual level $\varepsilon'_{0}=\varepsilon_{0}+\Re\Sigma_{0}$. Let us consider the other special point, i.e. $P$, where $\vec{\Xi}=\vec{\Sigma}_{1}$ or $\vec{z}=2\vec{\Sigma}_{1}$. Therefore, this point can be regarded as an extended point from the zero $A$ case since the same singularity at the Fermi level is preserved (except for an enhancement factor 2). The picture is that introducing the $A$-function doubles the effect of the interaction  and it is then easier to observe the Kondo resonance compared with  the zero $A$ case.

So far, we discussed the meaning of the constraint condition in Eq. (\ref{singularcondition1}). Now let's discuss the meaning of Eq. (\ref{singularcondition}). If we define $\Sigma_{0}=\rho_{0}e^{i\theta}$, we get
\begin{equation}
\vec{\Xi}'=\vec{\Xi}e^{-2i\theta}=\vec{\Sigma}_{1},
\label{xip}
\end{equation}
where $\vec{\Xi}'$, $\vec{\Xi}$ and $\vec{\Sigma}_{1}$ are viewed as vectors in the complex plane. Then the meaning of Eq. (\ref{xip}) is that $\vec{\Xi}'$ is obtained by rotating the vector $\vec{\Xi}$ by an angle $-2\theta$ while maintaining its magnitude. After this rotation, the new vector $\vec{\Xi}'$ (the "P" point on the ring in Fig. \ref{circle}(b)) should be equal to the vector $\vec{\Sigma}_{1}$. This situation is shown in Fig. \ref{circle}(b). When the end of the vector $\vec{\Xi}$ is not lying on the ring for $\omega=\mu$, say beyond it as shown in Fig. \ref{circle}(c), but the rotated vector $\vec{\Xi}'$ has the same real part of $\vec{\Sigma}_{1}$, we may observe a Kondo resonance.

\section{Summary}
In this work, we showed how the Kondo singlet develops for a magnetic adatom on graphene. We started from an Anderson model and kept the higher order contributions in the hybridization. Therefore, in this model, charge fluctuations are taken into account already.
Analytical Green's function is derived under a standard decoupling scheme. The Kondo resonance can thus be studied explicitly.
It is found that the Kondo resonance takes place in a much narrow range for the impurity level with respect to the Fermi energy which can be tuned by the gate voltage. We show that this parameter window is proportional to the Fermi energy $|\mu|$ and is well  narrower than that for the
 case of a normal metal host.
Numerical  calculations for LDOS  substantiate this analysis. The singularity in the full Green function is also analyzed in a transparent geometrical method, i.e. a singular ring in the complex plane. The relation between the various selfenergies is revealed.

\begin{acknowledgments}{ We thank Vitalii Dugaev for interesting discussions.
This work was supported by the DFG and  the state Saxony-Anhalt, Germany. }
\end{acknowledgments}

\appendix
\section{The calculations of the selfenergy $\Sigma_{1\bar{\sigma}}$}
\begin{figure}[tbh]
\includegraphics[width=0.35\textwidth]{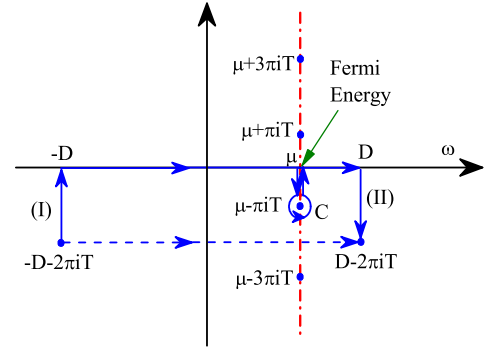}
\caption{(color online) The contour loop for the calculation of the selfenergy $\Sigma_{1\bar{\sigma}}$ is shown. The dashed path is deformed into the path indicated by solid lines with a circle C. \label{contour}}
\end{figure}

We introduce
\begin{equation}
\Phi^{\pm}(\omega)=\int_{-D}^{D}\frac{d\epsilon|\epsilon|f(\epsilon)}{\omega^{\pm}-\epsilon},
\label{phidefine}
\end{equation}
where $\omega^{\pm}=\omega\pm0^{+}$. To calculate it, we define
\begin{equation}
\chi^{\pm}(\omega)=\int_{-D}^{D}\frac{d\epsilon|\epsilon|[f(\epsilon)-1/2]}{\omega^{\pm}-\epsilon}.
\label{chidefine}
\end{equation}
The we make use of the relation
\begin{equation}
\Phi^{\pm}(\omega)=\chi^{\pm}(\omega)+(2\eta)^{-1}\Sigma_{0}(\omega^{\pm}).
\label{phichi}
\end{equation}
The calculation of $\chi$ are performed on an analytical contour which  is shown in Fig. \ref{contour}.

Firstly, we need to clarify the meaning of $|\epsilon|$ in an analytical contour sense. From some further analysis we infer $|\Re(\epsilon)|$ when $\epsilon$ is complex. The Fermi-Dirac distribution function has periodicity on the  imaginary axis through $2\pi iT$ (we set  the Boltzmann factor $k_{B}=1$), where $T$ is the temperature. Therefore, we find
\begin{equation}
\chi^{\pm}(\omega+2\pi iT)=\int_{-D-2\pi iT}^{D-2\pi iT}\frac{d\omega''|\omega''|\left[f(\omega'')-\frac{1}{2}\right]}{\omega-\omega''+i0^{+}},
\label{chishift}
\end{equation}
when we shift the argument by $2\pi iT$. The integral path in Eq. (\ref{chishift}) can be deformed into $-D-2\pi iT\mapsto-D\mapsto D\mapsto D-2\pi iT+\text{pole C}$. The pole is sitting at $\mu-\pi iT$. Therefore, we get
\begin{equation}
\chi^{+}(\omega+2\pi iT)=\chi^{+}(\omega)+\oint_{\text{C}}+(\text{I})+(\text{II}),
\label{chicontour}
\end{equation}
where the second term is a circular integral around the pole along the path C, the third and the fourth terms are integrals along the path (I) and (II), respectively. The second term can be obtained by using the residue calculus as
\begin{equation}
\oint_{\text{C}}=-|\mu|\frac{2\pi iT}{(\omega-\mu)+\pi iT}.
\label{loopc}
\end{equation}
The third and the fourth terms can be calculated as
\begin{eqnarray}
&&(\text{I})+(\text{II})= \label{pathIandII}\\
&&\frac{|D|}{2}\ln\left(\frac{(\omega-D+2\pi iT+i0^{+})(\omega+D+2\pi iT+i0^{+})}{(\omega-D+i0^{+})(\omega+D+i0^{+})}\right).\notag
\end{eqnarray}
Thus, we obtain the relation
\begin{eqnarray}
&-&\frac{\Phi^{+}(\omega+2\pi iT)}{|\mu|}+\frac{|D|}{2|\mu|} \notag\\
&\times& \ln\left(\frac{(\omega-D+2\pi iT+i0^{+})(\omega+D+2\pi iT+i0^{+})}{(2\pi iT)^{2}}\right) \notag \\
&=&-\frac{\Phi^{+}(\omega)}{|\mu|}+\frac{|D|}{2|\mu|}\ln\left(\frac{(\omega-D+i0^{+})(\omega+D+i0^{+})}{(2\pi iT)^{2}}\right) \notag\\
&+&\frac{2\pi iT}{(\omega-\mu)+\pi iT}.
\label{relation}
\end{eqnarray}
Defining $z=\frac{1}{2}+\frac{\omega-\mu}{2\pi iT}$, Eq.(\ref{relation}) recovers a relation of the digamma function
\begin{equation}
\psi(z+1)=\psi(z)+\frac{1}{z},
\label{digamma}
\end{equation}
leading to
\begin{eqnarray}
\chi^{\pm}(\omega) &=& \frac{|D|}{2}\ln\left(\frac{|\omega^{2}-D^{2}|}{(2\pi T)^{2}}\right)-|\mu|\psi\left(z^{\pm}\right), \notag\\
z^{\pm} &=& \frac{1}{2}\pm\frac{\omega-\mu}{2\pi iT}.
\label{chifianl}
\end{eqnarray}
From the definition of $\Sigma_{1\bar{\sigma}}(\omega^{\pm})$, we obtain
\begin{equation}
\Sigma_{1\bar{\sigma}}(\omega^{\pm})=\eta\Phi^{\pm}(\omega).
\label{self1psi}
\end{equation}
Substituting the corresponding expressions, we find  Eq. (\ref{self12}).



\begin{thebibliography}{99}

\bibitem{graphene} K. S. Novoselov, A. K. Geim, S. V. Morozov, D. Jiang, Y. Zhang, S. V. Dubonos, I. V. Grigorieva, and A. A. Firsov,
Science {\bf306}, 666 (2004); Y. Zhang, J. P. Small, M. E. S. Amori,
and P. Kim, Phys. Rev. Lett. {\bf94}, 176803 (2005); C. Berger, Z. Song, T. Li, X. Li, A. Y.
Ogbazghi, R. Feng, Z. Dai, A. N. Marchenkov,  E. H. Conrad, P. N.
First, and W. A. de Heer, J. Phys. Chem. B {\bf108}, 19912 (2004); V. P. Gusynin, S. G. Sharapov, and J. P. Carbotte, Inter. J. Mod. Phys. B, \textbf{21}, 4611 (2007); A. H. Castro Neto, F. Guinea, N. M. R. Peres, K. S. Novoselov, and A. K. Geim, Rev. Mod. Phys. \textbf{81}, 109 (2009).

\bibitem{grapheneold} P. R. Wallace, Phys. Rev. \textbf{71}, 622 (1947); W. M. Lomer, Proc. Roy. Soc. London A, \textbf{227}, 330 (1955);  J. McClure, Phys. Rev. \textbf{104}, 666 (1956); \textbf{108}, 612 (1957); J. C. Slonczewski, and P. R. Weiss, ibid. \textbf{109}, 272 (1958); F. Bassani, and G. P. Parravicini, IL Nuovo Cimento B, \textbf{50}, 95 (1967).


\bibitem{meyer}  J. C. Meyer, C. O. Girit, M. F. Crommie, and A. Zettl, Nature \textbf{454}, 319 (2008).

\bibitem{cornaglia} P. S. Cornaglia, G. Usaj, and C. A. Balseiro, Phys.
Rev. Lett. \textbf{102}, 046801 (2009).

\bibitem{zhuang} H. -B. Zhuang, Q. F. Sun, and X. C. Xie, Europhys. Lett. \textbf{86}, 58004 (2009).

\bibitem{chan} K. T. Chan, H. Lee, and M. L. Cohen, Phys. Rev. B \textbf{83}, 035405 (2011).

\bibitem{kotov} V. N. Kotov, B. Uchoa, V. M. Pereira, A. H. Castro Neto, and F. Guinea, arXiv: cond-matt 1012.3484.

\bibitem{uchoa} B. Uchoa, V. N. Kotov, N. M. R. Peres, and A. H. Castro Neto, Phys. Rev. Lett. {\bf101}, 026805 (2008).

\bibitem{ding} K. -H. Ding, Z. -G. Zhu, J. Berakdar, J. Phys.: Condens. Matter \textbf{21}, 182002 (2009).


\bibitem{uchoa0906} B. Uchoa, L. Yang, S. -W. Tsai, N. M. R. Peres, and A. H. Castro Neto, Phys. Rev. Lett. \textbf{103}, 206804 (2009).


\bibitem{hentschel} M. Hentschel, and F. Guinea, Phys. Rev. B \textbf{76}, 115407 (2007).

\bibitem{dora} B. D\'{o}ra, and P. Thalmeier, hys. Rev. B \textbf{76}, 115435 (2007).

\bibitem{sengupta} K. Sengupta, and G. Baskaran, Phys. Rev. B {\bf77}, 045417 (2008).

\bibitem{jacob} D. Jacob, and G. Kotliar, Phys. Rev. B \textbf{82}, 085423 (2010).

\bibitem{dellanna} L. Dell'Anna, J. Stat. Mech. P01007 (2010).

\bibitem{vojta} M. Vojta, L. Fritz, and R. Bulla, Europhys. Lett. \textbf{90}, 27006 (2010).

\bibitem{zhu10} Zhen-Gang Zhu, Kai-He Ding, and Jamal Berakdar, Europhys. Lett. \textbf{90}, 67001 (2010).

\bibitem{zhu2011} Zhen-Gang Zhu, and Jamal Berakdar, Phys. Rev. B \textbf{83},  195404 (2011).

\bibitem{uchoa11} B. Uchoa, T. G. Rappoport, and A. H. Castro Neto, Phys. Rev. Lett. \textbf{106}, 016801 (2011).

\bibitem{fradkin} D. Withoff, and E. Fradkin, Phys. Rev. Lett. \textbf{64}, 1835 (1990); C. R. Cassanello, and E. Fradkin, Phys. Rev. B \textbf{53},  15079 (1996).

\bibitem{buxton} C. Gonzalez-Buxton, and K. Ingersent, Phys. Rev. B \textbf{57},  14254 (1998).

\bibitem{fritz} M. Vojta, and L. Fritz, Phys. Rev. B \textbf{70}, 094502 (2004); L. Fritz, and M. Vojta, Phys. Rev. B \textbf{70}, 214427 (2004).

\bibitem{bulla} R. Bulla, T. A. Costi, and T. Pruschke, Rev. Mod. Phys. \textbf{80}, 395 (2008).


\bibitem{lacroix} C. Lacroix, J. Phys. F: Metal Phys. \textbf{11},
2389 (1981).

\bibitem{zubarev} D. N. Zubarev, Sov. Phys. Usp. \textbf{3}, 320 (1960).

\bibitem{gf} Zhen-Gang Zhu, Phys. Lett. A \textbf{372}, 695 (2008); Q. -F. Sun, and H. Guo, Phys. Rev. B \textbf{66}, 155308 (2002); N. Sergueev, Q. -F. Sun, H. Guo, B. G. Wang, and J. Wang, Phys. Rev. B \textbf{65}, 165303 (2002); B. Dong, H. L. Cui, S. Y. Liu, and X. L, Lei, J. Phys.: Condens. Matter \textbf{15}, 8435 (2003).

\bibitem{zhanggm} G. -M. Zhang, H. Hu, and L. Yu, Phys. Rev. Lett. \textbf{86},704 (2001).

\bibitem{hewson} A. C. Hewson, \textit{The Kondo problem to heavy
fermions}, (Cambridge Uni. Press, 1993).

\bibitem{theumann} A. Theumann, Phys. Rev. \textbf{178}, 978
(1969).

\bibitem{appelbaum} J. A. Appelbaum, and D. R. Penn, Phys. Rev. \textbf{188}, 874 (1969).

\bibitem{eom} Y. Meir, N. S. Wingreen, and P. A. Lee, Phys. Rev. Lett. \textbf{70}, 2601 (1993); V. Kashcheyevs, A. Aharony, and O. Entin-Wohlman, Phys. Rev. B \textbf{73}, 125338 (2006).

\bibitem{haldane} F. D. M. Haldane, Phys. Rev. Lett. \textbf{40}, 416 (1978).


%
%
%
%






\end{thebibliography}
\end{document}